\documentclass{tMOP2e}
\usepackage{graphicx}% eps files
\usepackage{bm}% bold math
\usepackage{amsmath}% equation formatting
\usepackage{amssymb}% math symbols
\newlength{\DL}

\newlength{\pwd} 
\begin{document} 
%******************************************* 
\title{Real-time quantum trajectories for classically allowed dynamics in strong laser fields} 
%*******************************************
\author{
\name{L.\ I.\ Plimak\textsuperscript{a}$^{\ast}$\thanks{$^\ast$Corresponding author. Email: lev.plimak@mbi-berlin.de}  
and Misha Yu.\ Ivanov\textsuperscript{a,b,c,d}% 
%and M. Frolov\textsuperscript{d}
}
\affil{\textsuperscript{a}Max Born Institute for Nonlinear Optics and Short Pulse Spectroscopy, 12489 Berlin, Germany}
\affil{\textsuperscript{b}Department of Physics, Imperial College London, South Kensington Campus, SW7 2AZ} 
\affil{\textsuperscript{c}Department of Physics, Humboldt University, Newtonstr. 15, 12489 Berlin, Germany} 
\affil{\textsuperscript{d}Department of Physics, Voronezh State University, Universitetskaya pl. 1, Voronezh,  Russia, 394036} 
}
%*******************************************
\maketitle 
%*******************************************
\begin{abstract} 
%*******************************************
Both the physical picture of the dynamics of atoms and molecules in intense infrared fields and its theoretical description use the concept of electron trajectories. Here we address a key question which arises in this context: Are distinctly quantum features of these trajectories, such as the complex-valued coordinates, physically relevant in the classically allowed region of phase space, and what is their origin? First, we argue that solutions of classical equations of motion can account for quantum effects. To this end, we construct an exact solution to the classical Hamilton-Jacobi equation\ which accounts for dynamics of the wave packet, and show that this solution is physically correct in the limit \mbox{$\hbar \to 0$}. Second, we show that imaginary components of classical trajectories are directly linked to the finite 
size of the initial wavepacket in momentum space. This way, if the electronic wavepacket produced by optical tunneling in strong infrared fiels is localised both in coordinate and momentum, its motion after tunneling {\em ipso facto\/} cannot be described with purely classical trajectories -- in contrast to popular models in the literature. 
%******************************************* 
\end{abstract}
%******************************************* 
\section{Introduction} 
%********************************************************
Nonclassical solutions to the Hamilton-Jacobi equation\ (HJE) \cite{LLI}, known as quantum trajectories, play an important role in strong-field physics \cite{Salieres2001Sc,Mairesse2003Sc,Dudovich2012N,KrauszIvanov}. They allow one to seamlessly incorporate quantum aspects of the dynamics, notably the laser-induced tunnelling, keeping at the same time a physical insight into the problem. 

The assumption underlying many theoretical models of strong-field ionization dynamics -- from the famous three-step model \cite{Corkum1993,CorkumKrausz2007} to Coulomb-corrected models of strong-field ionization \cite{Popruzhenko2008PRA1,Popruzhenko2008JMO,Popruzhenko2008PRA2,Popruzhenko2010PRL} to the analysis of attoclock experiments \cite{Keller2011NP,Keller2012PRL,Keller2013NJP,Keller2013PRL} -- is that quantum trajectories are limited to the motion in imaginary time in the classically forbidden region during the tunnelling stage. So far as the electron has emerged from under the barrier, its motion to the detector is expected to be classical in the true sense of the term. Consequently, in simulations, both the coordinate and the momentum characteristic of its trajectory are presumed to be real \cite{Keller2011NP,Keller2012PRL,Keller2013NJP,Keller2013PRL}. 

The notable exception are the studies of high harmonic generation and correlated double ionisation, where the electron momentum remains complex after ionisation, until its recollision with the parent ion (see, e.g., \cite{Becker2014} and references therein). Furthermore, it has recently been shown \cite{OlgaARM1,OlgaARM2,OlgaARM3,OlgaARM4,OlgaARM5} that the presence of an imaginary component in an otherwise classically-behaving trajectory after tunnelling is crucial for quantitative agreement with ab-initio calculations even in the absense of recollision \cite{Attoclock}. 
While for short-range potentials imaginary part of the trajectory coordinate remains hidden from view, as it does not impact ionization yields or photoelectron spectra, for long-range potentials it leads to qualitative and quantitative consequences \cite{SmirnovaTorlina}. Counter-intuitive as they may seem, quantum trajectories appear to exist in any case of classically permitted motion, be it from tunneling to recollision or from tunneling to the detector. 

We argue that resolution of this controversy is in the fact that solutions to classical equations of motion, such as the HJE, need not be by themselves classical. They can account to some extent for quantum effects. Emergence of quantum trajectories in classically permitted regions is thus associated with a certain hierarchy of quantum effects, with some accounted for in solutions to the HJE, and others neglected. 

In this note, we demonstrate real-time quantum trajectories for the simplest textbook example: the motion of a Gaussian wavepacket in a field of homogeneous time-dependent force. The assumption of a Gaussian wavepacket is typical, e.g., in the analyses of the electron dynamics after tunneling induced by the infrared laser pulses. Following the standard logic of separation of ``exponential terms'' from ``pre-exponential factors'', we regard normalisation of the wavepacket ``pre-exponential''. We show that this is justified in the semiclassical limit \mbox{$\hbar \to 0$}. The ``exponential terms'' then add up to an {\em exact nonclassical\/} solution to the HJE. Quantum interference (phase) effects, which are, according to such logic, ``exponential'', are by construction included in this solution. Nonclassical nature (complexity) of the latter is shown to be associated with localisation of the Gaussian packet in coordinate and momentum. 

%******************************************* 
\section{WKB Ansatz revisited} 
%******************************************* 
\subsection{WKB basics} 
%********************************************************
Formally, the aforementioned exact nonclassical solution to the HJE emerges within a quasiclassical (WKB) approximation \cite{WKB-W,WKB-K,WKB-B} to evolution of a Gaussian wave packet under the influence of a homogeneous force. 
Under the said approximation, one solves the {Schr\"odinger}\ equation, 
%=============================================
\begin{align} 
i\hbar \partial_t\psi ({\mbox{\rm\boldmath$r$}},t) = \protect\Big [ -\frac{\hbar ^2}{2m}{\partial_{\mbox{\rm\scriptsize\boldmath$r$}_{}}^2} &+V({\mbox{\rm\boldmath$r$}},t) \Big ]\psi ({\mbox{\rm\boldmath$r$}},t) , 
%}{%
\label{eq:88UL} % \nonumber % \Z 
\end{align}%
%+++++++++++++++++++++++++++++++++++++++++++++
by making the Ansatz, 
%=============================================
\begin{align} 
\psi ({\mbox{\rm\boldmath$r$}},t) = \exp
\protect\bigg [ \frac{iS({\mbox{\rm\boldmath$r$}},t)}{\hbar }+i\sigma ({\mbox{\rm\boldmath$r$}},t) \bigg ] . 
%}{%
\label{eq:59AA} % \nonumber % \Z 
\end{align}%
%+++++++++++++++++++++++++++++++++++++++++++++
The ``WKB action'' \mbox{$S({\mbox{\rm\boldmath$r$}},t)$} obeys the equation, 
%=============================================
\begin{align} 
-\partial_{t} S({\mbox{\rm\boldmath$r$}},t)= \frac{1}{2m}
\protect [ \partial_{\mbox{\rm\scriptsize\boldmath$r$}_{}}S({\mbox{\rm\boldmath$r$}},t) ]^2 + V({\mbox{\rm\boldmath$r$}},t) , 
%}{%
\label{eq:61AC} % \nonumber % \Z 
\end{align}%
%+++++++++++++++++++++++++++++++++++++++++++++
which is a classical HJE for the Hamiltonian,
%=============================================
\begin{align} 
H({\mbox{\rm\boldmath$p$}},{\mbox{\rm\boldmath$r$}},t) = \frac{{\mbox{\rm\boldmath$p$}}^2}{2m} + V({\mbox{\rm\boldmath$r$}},t).
%}{%
\label{eq:90DC} % \nonumber % \Z 
\end{align}%
%+++++++++++++++++++++++++++++++++++++++++++++
Substituting this to (\ref{eq:88UL}) we find the equation for \mbox{$\sigma ({\mbox{\rm\boldmath$r$}},t)$}, 
%=============================================
\begin{align} 
-\partial_{t}\sigma({\mbox{\rm\boldmath$r$}},t) &= \frac{1}{m}\partial_{\mbox{\rm\scriptsize\boldmath$r$}_{}}S({\mbox{\rm\boldmath$r$}},t)\partial_{\mbox{\rm\scriptsize\boldmath$r$}_{}}\sigma({\mbox{\rm\boldmath$r$}},t)- \frac{i}{2m}\Delta S({\mbox{\rm\boldmath$r$}},t) 
%\np 
+ \frac{\hbar }{2m}
\protect \{ \protect [ \partial_{\mbox{\rm\scriptsize\boldmath$r$}_{}}\sigma ({\mbox{\rm\boldmath$r$}},t) ]^2 - i\Delta\sigma ({\mbox{\rm\boldmath$r$}},t) \} . 
%}{%
\label{eq:60AB} % \nonumber % \Z 
\end{align}%
%+++++++++++++++++++++++++++++++++++++++++++++
In the quasiclassical (WKB) approximation, the term proportional to \mbox{$\hbar $} in (\ref{eq:60AB}) is dropped. 

%********************************************************
\subsection{Is WKB action necessarily classical?}%
\label{ch:38QK}
%********************************************************
\mbox{Equations (\ref{eq:61AC})}, (\ref{eq:60AB}) follow by substituting the WKB Ansatz (\ref{eq:59AA}) in the {Schr\"odinger}\ equation (\ref{eq:88UL}) and grouping the terms according to powers of \mbox{$\hbar $}. However, this is no more than a leading consideration. By themselves, \mbox{Eqs.\ (\ref{eq:61AC})}, (\ref{eq:60AB}) are exact. More precizely speaking, with the HJE (\ref{eq:61AC}) postulated, \mbox{Eqs.\ (\ref{eq:59AA})} and (\ref{eq:60AB}) are equivalent. 

There is nothing formally wrong with accommodating quantum effects in \mbox{$S({\mbox{\rm\boldmath$r$}},t)$}. 
Such approach may be beneficial if a particular quantum effect exceeds in magnitude certain classical ones. 
Indeed, think of a wave packet propagating in a potential. Such quantum motion can always be separated into the classical (trajectorial) dynamics and evolution of the wave-packet. If the potential is linear (a homogeneous force), the packet dynamics reduces to spread. For a potential that is only approximately linear (a nearly homogeneous force), the quantum spread may well happen to remain large compared to corrections to the classical trajectory. In this case, classification of effects according to their classical or quantum nature does not make much sense.

%********************************************************
\subsection{WKB action {\em versus\/} the classical action}%
\label{ch:76LF}
%********************************************************
Now, where is the {\em mathematical opening\/} for \mbox{$S({\mbox{\rm\boldmath$r$}},t)$} to include quantum effects? It is instructive to compare \mbox{$S({\mbox{\rm\boldmath$r$}},t)$} to the classical action proper, 
%=============================================
\begin{gather} 
\begin{gathered} 
S({\mbox{\rm\boldmath$r$}},t;{\mbox{\rm\boldmath$r$}}_0,t_0) = \int_{t_0}^{t} 
L({\mbox{\rm\boldmath$p$}}(t'),{\mbox{\rm\boldmath$r$}}(t'),t' ) dt' , 
\end{gathered} 
\label{eq:75LE} % \nonumber % \Z 
\end{gather}%
%+++++++++++++++++++++++++++++++++++++++++++++
where the classical trajectory \mbox{${\mbox{\rm\boldmath$p$}}(t),{\mbox{\rm\boldmath$r$}}(t)$} is specified by the conditions, 
%=============================================
\begin{gather} 
\begin{gathered} 
{\mbox{\rm\boldmath$r$}}(t) = {\mbox{\rm\boldmath$r$}}, \quad {\mbox{\rm\boldmath$r$}}(t_0) = {\mbox{\rm\boldmath$r$}}_0 . 
\end{gathered} 
\label{eq:84NB} % \nonumber % \Z 
\end{gather}%
%+++++++++++++++++++++++++++++++++++++++++++++
The classical action is a ``two-ended'' quantity. It is specified by {\em two boundary conditions\/} for trajectories, and obeys in fact {\em two HJE's\/}, Eq.\ (\ref{eq:61AC}) and the other one with \mbox{$\partial_t\to -\partial_{t_0}$}, \mbox{$\partial_{\mbox{\rm\scriptsize\boldmath$r$}_{}}\to -\partial_{\mbox{\rm\scriptsize\boldmath$r$}_{0}}$}, \mbox{$V({\mbox{\rm\boldmath$r$}},t)\to V({\mbox{\rm\boldmath$r$}}_0,t_0)$}. The WKB action \mbox{$S({\mbox{\rm\boldmath$r$}},t)$} is a ``one-ended'' quantity. It is specified by {\em one boundary condition\/}, \mbox{${\mbox{\rm\boldmath$r$}}(t)={\mbox{\rm\boldmath$r$}}$}, and obeys {\em one HJE\/} (\ref{eq:61AC}). That is, conditions specifying the WKB action are weaker than those for the classical action in the true meaning of the term. 

To see how this opening may work, consider restoration of quantum trajectories from a given WKB action. They are found from the equation, 
%=============================================
\begin{align} 
m\partial_{t} {\mbox{\rm\boldmath$r$}}= {\mbox{\rm\boldmath$p$}}({\mbox{\rm\boldmath$r$}},t) , 
%}{%
\label{eq:86CY} % \nonumber % \Z 
\end{align}%
%+++++++++++++++++++++++++++++++++++++++++++++
where \mbox{${\mbox{\rm\boldmath$p$}}({\mbox{\rm\boldmath$r$}},t)$} is the momentum as a function of \mbox{${\mbox{\rm\boldmath$r$}},t$}, 
%=============================================
\begin{align} 
{\mbox{\rm\boldmath$p$}}({\mbox{\rm\boldmath$r$}},t) = \partial_{\mbox{\rm\scriptsize\boldmath$r$}_{}}S({\mbox{\rm\boldmath$r$}},t). 
%}{%
\label{eq:70KZ} % \nonumber % \Z 
\end{align}%
%+++++++++++++++++++++++++++++++++++++++++++++
\mbox{Equation (\ref{eq:86CY})} is a system of three first-order differential equations; their solution depends on three integration constants. This fact is of utter importance. It shows that by specifying action one reduces the number of available integration constants from six -- cf.\ \mbox{Eq.\ (\ref{eq:84NB})} -- to three, thus imposing a condition on underlying trajectories. For the classical action, this is the boundary condition \mbox{${\mbox{\rm\boldmath$r$}}(t_0)={\mbox{\rm\boldmath$r$}}_0$}. However, in general, there is no guarantie that this condition may be interpreted classically. In turn, this would mean that the WKB action is itself a nonclassical quantity -- the fact that it obeys the HJE notwithstanding. 

%********************************************************
\section{Nonclassical solutions to the Hamilton-Jacobi equation}\label{ch:18NA}%
%********************************************************
\subsection{Gaussian wave packet in a field of homogeneous force}\label{ch:37QJ}%
%********************************************************
A physically meaningful example of a nonclassical WKB action emerges if considering evolution of a Gaussian wave packet in an arbitrary nonstationary homogeneous force field, 
%=============================================
\begin{align} 
V({\mbox{\rm\boldmath$r$}},t) = V_{\mathrm{L}}({\mbox{\rm\boldmath$r$}},t) = - {\mbox{\rm\boldmath$r$}}{\mbox{\rm\boldmath$F$}}(t) . 
%}{%
\label{eq:79CR} % \nonumber % \Z 
\end{align}%
%+++++++++++++++++++++++++++++++++++++++++++++
An exact solution to this problem reads, 
%=============================================
\begin{align} 
\psi ({\mbox{\rm\boldmath$r$}},t) &= \protect\bigg [ \frac{\Re D(t_0)}{\pi D^2(t)} \bigg ]^{3/4}
\exp\bigg(\frac{i}{\hbar }
\protect\Big \{ W_{\mathrm{L}}(t)+\protect [ {\mbox{\rm\boldmath$r$}}-{\mbox{\rm\boldmath$X$}}(t) ]{\mbox{\rm\boldmath$P$}}(t) \Big \} %\npp 
-\frac{
\protect [ {\mbox{\rm\boldmath$r$}}-{\mbox{\rm\boldmath$X$}}(t) ]^2 }{2D(t)} \bigg) . 
%}{%
\label{eq:64BZ} % \nonumber % \Z 
\end{align}%
%+++++++++++++++++++++++++++++++++++++++++++++
Here 
%=============================================
\begin{gather} 
\begin{gathered} 
D(t) = D(t_0) + \frac{i\hbar (t-t_0)}{m} 
\end{gathered} 
\label{eq:56XJ} % \nonumber % \Z 
\end{gather}%
%+++++++++++++++++++++++++++++++++++++++++++++
determines the width of the wave packet. 
Other quantities in (\ref{eq:64BZ}) are defined for a classical motion in potential (\ref{eq:79CR}): 
\mbox{${\mbox{\rm\boldmath$X$}}(t),{\mbox{\rm\boldmath$P$}}(t)$} is a classical trajectory obeying the Hamilton equations, 
%=============================================
\begin{align} 
\partial_t{\mbox{\rm\boldmath$X$}}(t) = \frac{{\mbox{\rm\boldmath$P$}}(t)}{m}, \quad \partial_t{\mbox{\rm\boldmath$P$}}(t) = {\mbox{\rm\boldmath$F$}}(t), 
%}{%
\label{eq:95DJ} % \nonumber % \Z 
\end{align}%
%+++++++++++++++++++++++++++++++++++++++++++++
and 
%=============================================
\begin{align} 
W_{\mathrm{L}}(t) = {\mbox{\rm\boldmath$X$}}(t){\mbox{\rm\boldmath$P$}}(t) - \int _{t_0}^{t}dt' \frac{{\mbox{\rm\boldmath$P$}}^2(t')}{2m} . 
%}{%
\label{eq:92DE} % \nonumber % \Z 
\end{align}%
%+++++++++++++++++++++++++++++++++++++++++++++
The intial conditions \mbox{${\mbox{\rm\boldmath$X$}}(t_0),\ {\mbox{\rm\boldmath$P$}}(t_0)$} and \mbox{$D(t_0)$} may be arbitrary, except the real part of \mbox{$D(t_0)$} must be positive to assure normalisation of (\ref{eq:64BZ}), 
%=============================================
\begin{align} 
\Re D(t_0)> 0 . 
%}{%
\label{eq:93DF} % \nonumber % \Z 
\end{align}%
%+++++++++++++++++++++++++++++++++++++++++++++
That (\ref{eq:64BZ}) solves the {Schr\"odinger}\ equation (\ref{eq:88UL}) with \mbox{$V=V_{\mathrm{L}}$} is readily verified by direct calculation. 

%********************************************************
\subsection{Nonclassical solution to the HJE}\label{ch:08MQ}%
%********************************************************
Solution (\ref{eq:64BZ}) should be compared to the WKB Ansatz (\ref{eq:59AA}). The obvious choice of the WKB action \mbox{$S({\mbox{\rm\boldmath$r$}},t)$} is to identify it with the combination in curly brackets. A less obvious choice is to account in \mbox{$S({\mbox{\rm\boldmath$r$}},t)$} for the {\em quantum phase stucture\/} this packet acquires when spreading, 
%=============================================
\begin{align} 
S({\mbox{\rm\boldmath$r$}},t) = W_{\mathrm{L}}(t)+\protect [ {\mbox{\rm\boldmath$r$}}-{\mbox{\rm\boldmath$X$}}(t) ]{\mbox{\rm\boldmath$P$}}(t)+\frac{i\hbar 
\protect [ {\mbox{\rm\boldmath$r$}}-{\mbox{\rm\boldmath$X$}}(t) ]^2 }{2D(t)}. 
%}{%
\label{eq:67CC} % \nonumber % \Z 
\end{align}%
%+++++++++++++++++++++++++++++++++++++++++++++
The ``quantum correction'' is then limited to normalisation, 
%=============================================
\begin{align} 
i\sigma ({\mbox{\rm\boldmath$r$}},t) = i\sigma (t) = \frac{3}{4}\ln\frac{\Re D(t_0)}{\pi D^2(t)} . 
%}{%
\label{eq:68CD} % \nonumber % \Z 
\end{align}%
%+++++++++++++++++++++++++++++++++++++++++++++
To verify that (\ref{eq:67CC}) indeed obeys the HJE, 
%=============================================
\begin{align} 
\partial_{t} S({\mbox{\rm\boldmath$r$}},t)+ \frac{1}{2m}
\protect [ \partial_{\mbox{\rm\scriptsize\boldmath$r$}_{}}S({\mbox{\rm\boldmath$r$}},t) ]^2 -{\mbox{\rm\boldmath$r$}}{\mbox{\rm\boldmath$F$}}(t) = 0, 
%}{%
\label{eq:69CE} % \nonumber % \Z 
\end{align}%
%+++++++++++++++++++++++++++++++++++++++++++++
is again straightforward. 

%********************************************************
\subsection{Quantum trajectories}\label{ch:09MR}%
%********************************************************
So what kind of motion does \mbox{$S({\mbox{\rm\boldmath$r$}},t)$} describe? The momentum as a function of \mbox{${\mbox{\rm\boldmath$r$}},t$} reads, 
%=============================================
\begin{align} 
{\mbox{\rm\boldmath$p$}}({\mbox{\rm\boldmath$r$}},t) = {\mbox{\rm\boldmath$P$}}(t) +\frac{i\hbar 
\protect [ {\mbox{\rm\boldmath$r$}}-{\mbox{\rm\boldmath$X$}}(t) ] }{D(t)} . 
%}{%
\label{eq:70CF} % \nonumber % \Z 
\end{align}%
%+++++++++++++++++++++++++++++++++++++++++++++
Solving equation (\ref{eq:86CY}) 
we obtain the trajectories, 
%=============================================
\begin{align} 
{\mbox{\rm\boldmath$r$}}(t) = {\mbox{\rm\boldmath$X$}}(t) + \frac{D(t){\mbox{\rm\boldmath$p$}}_{\mathrm{i}}}{i\hbar }, \quad
{\mbox{\rm\boldmath$p$}}(t) = {\mbox{\rm\boldmath$P$}}(t) + {\mbox{\rm\boldmath$p$}}_{\mathrm{i}},
%}{%
\label{eq:87CZ} % \nonumber % \Z 
\end{align}%
%+++++++++++++++++++++++++++++++++++++++++++++
where 
%=============================================
\begin{align} 
{\mbox{\rm\boldmath$p$}}_{\mathrm{i}} = \frac{i\hbar 
\protect [ {\mbox{\rm\boldmath$r$}}(t_0)-{\mbox{\rm\boldmath$X$}}_0(t) ] }{D(t_0)}
%}{%
\label{eq:88DA} % \nonumber % \Z 
\end{align}%
%+++++++++++++++++++++++++++++++++++++++++++++
is the triad of integration constants we spoke about in \mbox{Sec.\ \ref{ch:76LF}}. 

While functions (\ref{eq:87CZ}) do obey classical equations of motion, they are not what one would normally call ``a classical trajectory''. To start with, they are complex. 
Reality of the momentum may be maintained (with real \mbox{${\mbox{\rm\boldmath$p$}}_{\mathrm{i}}$}), but not reality of the coordinate. These peculiarities are rooted in the initial condition, 
%=============================================
\begin{align} 
i\hbar \protect [ {\mbox{\rm\boldmath$r$}}(t_0)-{\mbox{\rm\boldmath$X$}}(t_0) ] = D(t_0) \protect [ {\mbox{\rm\boldmath$p$}}(t_0)-{\mbox{\rm\boldmath$P$}}(t_0) ] .
%}{%
\label{eq:89DB} % \nonumber % \Z 
\end{align}%
%+++++++++++++++++++++++++++++++++++++++++++++
It makes no sense in classical mechanics because of restriction (\ref{eq:93DF}). We stress that time $t$ remains real, so that the WKB action \mbox{$S({\mbox{\rm\boldmath$r$}},t)$} and the trajectory (\ref{eq:87CZ}) are characteristic of the {\em classically permitted\/} motion. 

%******************************************* 
\section{Discussion}\label{ch:07MP} 
%******************************************* 
The WKB action (\ref{eq:67CC}) emerged as an {\em ad hoc\/} quantity. In fact its choice may be supported by scaling arguments in the classical limit \mbox{$\hbar \to 0$}. Assume that \mbox{$D(t)$} is real at some time \mbox{$t_{\mathrm{in}}$}. For large propagation times, the wave packet spread becomes much larger than its initial width (cf.\ \mbox{Sec.\ \ref{ch:38QK}}). Neglecting the latter, 
%=============================================
\begin{align} 
D(t) \approx \frac{i\hbar (t-t_{\mathrm{in}})}{m}, 
%}{%
\label{eq:05MM} % \nonumber % \Z 
\end{align}%
%+++++++++++++++++++++++++++++++++++++++++++++
and 
%=============================================
\begin{align} 
S({\mbox{\rm\boldmath$r$}},t) \approx W_{\mathrm{L}}(t)+\protect [ {\mbox{\rm\boldmath$r$}}-{\mbox{\rm\boldmath$X$}}(t) ]{\mbox{\rm\boldmath$P$}}(t)+\frac{m 
\protect [ {\mbox{\rm\boldmath$r$}}-{\mbox{\rm\boldmath$X$}}(t) ]^2 }{2(t-t_{\mathrm{in}})}. 
%}{%
\label{eq:06MN} % \nonumber % \Z 
\end{align}%
%+++++++++++++++++++++++++++++++++++++++++++++
Planck's constant has dropped out. In the approximation (\ref{eq:05MM}), it only survives in the preexponential normalisation factor. Influence of this factor in the limit \mbox{$\hbar \to 0$} is small. This is clear from comparing \mbox{$S({\mbox{\rm\boldmath$r$}},t)/\hbar $} to \mbox{$\sigma ({\mbox{\rm\boldmath$r$}},t)$}. The former scales as \mbox{$1/\hbar $} while the latter as \mbox{$\ln\hbar $}. 

It is interesting and instructive to note that in the approximation (\ref{eq:05MM}) trajectories (\ref{eq:87CZ}) and the condition (\ref{eq:89DB}) become real, and start making formal sense in classical mechanics (with real \mbox{${\mbox{\rm\boldmath$p$}}_{\mathrm{i}}$}). In particular, \mbox{Eq.\ (\ref{eq:89DB})} becomes, 
%=============================================
\begin{align} 
{\mbox{\rm\boldmath$r$}}(t_0)-{\mbox{\rm\boldmath$X$}}(t_0) = \frac{t_0-t_{\mathrm{in}}}{m} \protect [ {\mbox{\rm\boldmath$p$}}(t_0)-{\mbox{\rm\boldmath$P$}}(t_0) ] .
%}{%
\label{eq:10MS} % \nonumber % \Z 
\end{align}%
%+++++++++++++++++++++++++++++++++++++++++++++
This condition makes perfect classical sense. Recalling \mbox{Eq.\ (\ref{eq:87CZ})} it may be rewritten as, 
%=============================================
\begin{align} 
{\mbox{\rm\boldmath$r$}}(t_0)-{\mbox{\rm\boldmath$X$}}(t_0) = \frac{(t_0-t_{\mathrm{in}}){\mbox{\rm\boldmath$p$}}_{\mathrm{i}}}{m} , 
%}{%
\label{eq:11MT} % \nonumber % \Z 
\end{align}%
%+++++++++++++++++++++++++++++++++++++++++++++
or simply 
%=============================================
\begin{align} 
{\mbox{\rm\boldmath$r$}}(t_{\mathrm{in}})={\mbox{\rm\boldmath$X$}}(t_{\mathrm{in}}). 
%}{%
\label{eq:41QN} % \nonumber % \Z 
\end{align}%
%+++++++++++++++++++++++++++++++++++++++++++++
That is, in the approximation (\ref{eq:05MM}), WKB action (\ref{eq:67CC}) reduces to the classical action (\ref{eq:75LE}), with \mbox{$t_0\to t_{\mathrm{in}}$} and \mbox{${\mbox{\rm\boldmath$r$}}_0\to {\mbox{\rm\boldmath$X$}}(t_{\mathrm{in}})$}. Up to a constant, 
%=============================================
\begin{align} 
S({\mbox{\rm\boldmath$r$}},t) = S({\mbox{\rm\boldmath$r$}},t;{\mbox{\rm\boldmath$X$}}(t_{\mathrm{in}}),t_{\mathrm{in}}) . 
%}{%
\label{eq:12MU} % \nonumber % \Z 
\end{align}%
%+++++++++++++++++++++++++++++++++++++++++++++
Physically, it describes a bunch of trajectories ``growing'' from a single point \mbox{${\mbox{\rm\boldmath$X$}}(t_{\mathrm{in}})$}, quite natural a result of neglecting the initial width of the packet. 

This way, nonclassically of the action (\ref{eq:67CC}) hinges on whether the initial width of the Gaussian pachet can or cannot be neglected. In any case, quantum mechanics prohibits \mbox{Eq.\ (\ref{eq:05MM})} as an exact relation, because it would render the wave packet unnormalisable. Complexity in (\ref{eq:87CZ}) and (\ref{eq:89DB}) are thus traces of quantum mechanics\ in ``nearly classical'' formulae. 

This is the case, for example, in interpretation of the attoclock experiments on tunneling. The deviation of the observed results from those based on classical modeling of the electron dynamics after ionization are attributed to the parameters of the initial Gaussian wavepacket assumed after tunneling. Once these parameters become important, the quantumness of the underlying trajectories has to be included. The quantitative consequences of this quantumness have been discussed in \cite{SmirnovaTorlina}.

In conclusion, a few words so as to put this paper in a wider perspective. After the initial shock -- how on earth the electron choses the state to leap to -- there has been a growing awareness that quantum and classical mechanics\ overlap rather than are mutually exclusive. Informally speaking, the quantum-classical border is an area rather than a line. This viewpoint has been especially fruitful in quantum optics, culminating in Glauber's photodetection theory \cite{GlauberPhDet,KelleyKleiner,GlauberTN} and phase-space techniques \cite{SchrCohSt,Feynman,BakerPhSp,Sudarshan,GlauberCohInc,CahillGlauber,AgarwalWolf,MandelWolf}. 

One of the defining features of the said overlap is that quantum processes may be governed by classical equations of motion. For the electromagnetic field, this extends in fact to full formal identity of the quantum and classical dynamics \cite{DirResp,Maxwell}. {\em Ipso facto\/}, such identity is impossible for equations of motion of matter, because it would mean rewriting the whole of quantum mechanics\ as a hidden-variable theory \cite{BellUnsp} (and this is not to mention fermions). 
The critical question is therefore, how far classical concepts actually extend into the realm of quantum mechanics. As is shown in this paper, one of the ``inhabitants'' of the quantum-classical overlap is the classical Hamilton-Jacobi equation. 

To summarise, the classical action is a solution to the HJE, but not {\em vice versa\/}. Even in real time, an arbitrary solution to the HJE needs not be classical. Specifying it implicitly imposes conditions on ``classical trajectories'', which may happen to be impossible for classical trajectories in the true meaning of the term. One may say that the Hamilton-Jacobi equation\ is in fact a {\em quantum equation\/}, of which some (but not all) solutions make sense in classical mechanics. ``Quantum'' solutions to the HJE are physically important within the quantum-classical overlap, i.e., in the limit \mbox{$\hbar \to 0$}. 
%******************************************* 
\section*{Acknowledgements} 
%******************************************* 
M.\ I.\ is grateful to the Voronezh State University, where part of this work
was performed, for generous hospitality. M.\ I.\ also acknowledges financial support of the EPSRC through the Program Grant on Attosecond Dynamics, Award No.\ EP/I032517, and of CORINF, a Marie Curie ITN of the European Union, Grant Agreement No. 264951.
%*******************************************
%\bibliographystyle{tMOP}\bibliography{QAction}\end{document} 
%*******************************************

%******************************************* 
\end{document}